\documentclass[
 aip, apl,
 amsmath,amssymb,
 reprint
]{revtex4-1}

\usepackage{hyperref}
\hypersetup{
  colorlinks=true,
  anchorcolor=blue,
  linkcolor=blue,
  citecolor=blue,
  filecolor=blue,
  urlcolor=blue,
}
\usepackage{graphicx}
\usepackage{bm}
\usepackage{color}  
\usepackage[separate-uncertainty=true,multi-part-units=single]{siunitx}

\begin{document}

\title{Wide-field diamond magnetometry with millihertz frequency resolution and nanotesla sensitivity}%

\newcommand{\affTitech}{School of Engineering, Department of Electrical and Electronic Engineering, Tokyo Institute of Technology, 2-12-1 Ookayama, Meguro-ku, Tokyo 152-8552, Japan}
\newcommand{\affPRESTO}{PRESTO, Japan Science and Technology Agency, Tokyo 102-0076, Japan}

\author{Kosuke Mizuno}
\author{Makoto Nakajima}
\affiliation{\affTitech}
\author{Hitoshi Ishiwata}
\affiliation{\affTitech}
\affiliation{\affPRESTO}
\author{Yuta Masuyama}
\author{Takayuki Iwasaki}
\author{Mutsuko Hatano}
\email[Author to whom correspondence should be addressed: ]{hatano.m.ab@m.titech.ac.jp}
\affiliation{\affTitech}

\date{\today}

\newcommand{\iQdyne}{\textit{i}Qdyne}
\newcommand{\SIrtHz}[2]{\SI{#1}{\mathrm{#2}/{\mathrm{Hz}}^{1/2}}}
\newcommand{\todo}[1]{\textcolor{red}{#1}}  

\begin{abstract}
The nitrogen-vacancy (NV) center in diamond allows room-temperature wide-field quantum magnetometry and metrology for a small volume, which is an important technology for applications in biology.
Although coherence of the NV center has a limited frequency resolution of diamond magnetometry to 10--\SI{100}{kHz}, recent studies have shown that a phase sensitive protocol can beat the coherence limit on a confocal setup.
Here, we report a new measurement protocol, ``{\iQdyne},'' for improving the frequency resolution of wide-field imaging beyond the coherence limit of the NV center.
We demonstrate wide-field magnetometry with a frequency resolution of \SI{238}{mHz} and a magnetic sensitivity of \SIrtHz{65}{nT}, which are superior to the conventional XY8-based technique, which paves the way to \textit{in vivo} microscale nuclear magnetic resonance imaging.
We find that the experimental performance of {\iQdyne} agrees well with that of an analytical model.
\end{abstract}

\maketitle

\begin{figure*}
  \includegraphics[width=17cm]{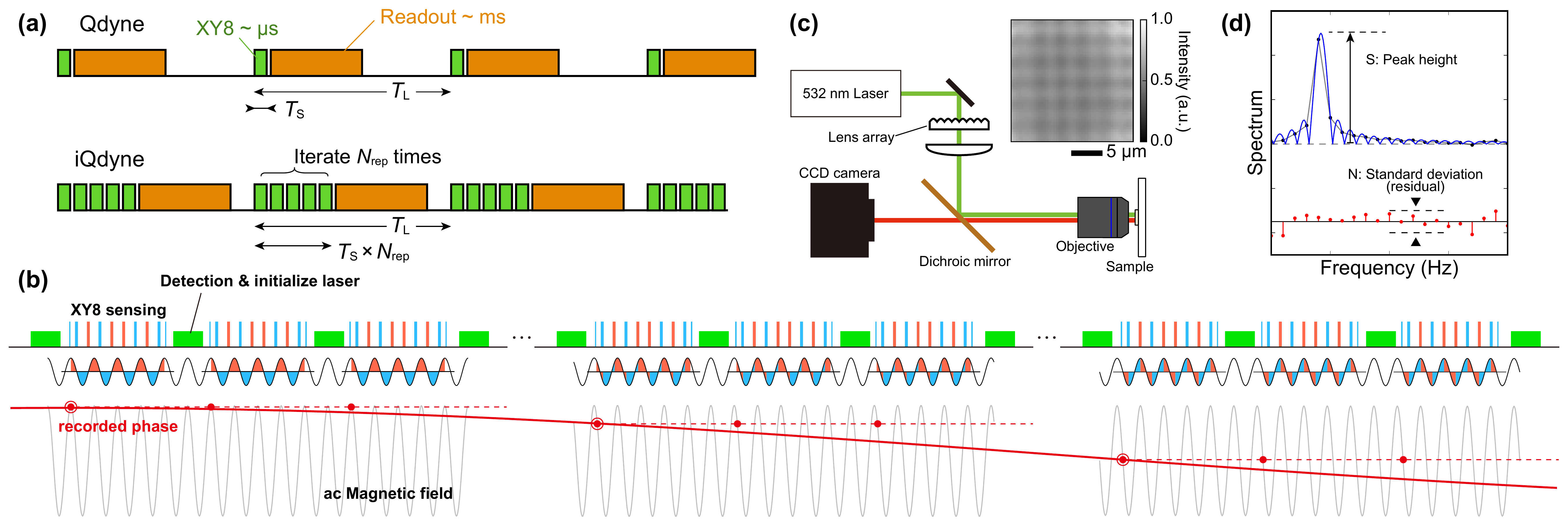}
  \caption{\label{fig:protocols} (a) Directly implemented Qdyne protocol with a slow detector such as a CCD camera and proposed ``iterative-Qdyne ({\iQdyne})'' protocol. The SNR is improved by iterating the XY8 measurement during an exposure time and adjusting the sampling interval to an integer multiple of the target period. (b) Principle of {\iQdyne} protocol. Each XY8 sequence in a certain measurement records the ac magnetic field (gray line) with an identical initial phase (red points), although the ac phase advance during the CCD readout. The red line is a guide for eye. (c) Schematic diagram of the wide-field optics. Inset: Wide-field distribution of the excitation intensity. (d) Definition of the SNR. ``S (N)'' corresponds to the peak height (standard deviation) in the Fourier spectrum.}
\end{figure*}
\begin{figure}
  \includegraphics[width=8.5cm]{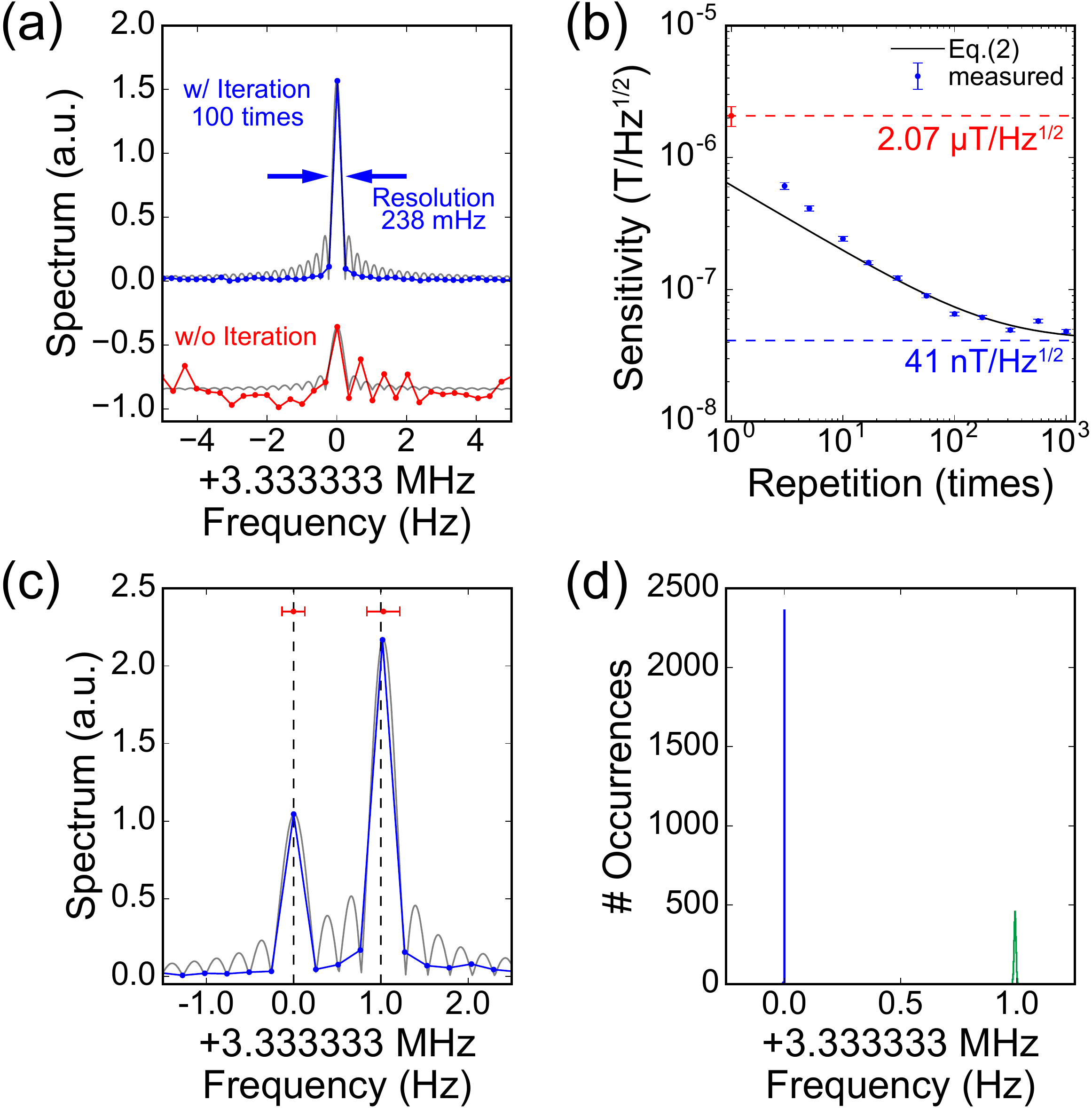}
  \caption{\label{fig:result_lockin} (a) Comparison of the directly implemented Qdyne (red) and {\iQdyne} protocols (blue). The solid line and dots represent obtained data, and gray curves represent the fitting function. (b) Sensitivity dependence on the iteration number of the measurement during an exposure. The red and blue dashed lines represent the directly implemented Qdyne sensitivity and optimal {\iQdyne} sensitivity $\eta_\infty$. The error bars are estimated by the standard deviation of the fitting parameters. (c) Result of {\iQdyne} with two different RF detection. The dashed black line, red points, and error bars represent the applied RF frequency, estimated frequency, and fitting precision, respectively. (d) Histogram of the estimated frequencies of each pixel.}
\end{figure}
The nitrogen-vacancy (NV) center~\cite{Taylor2008, Wrachtrup2016} composed of a substitutional nitrogen atom and a neighboring lattice vacancy has emerged as a breakthrough material for quantum sensing.\cite{Clevenson2015,Wolf2015,Barry2016,Pham2011,Glenn2017,Fu2017,Tetienne2017,Staudacher2013,Lovchinsky2016,Aslam2017} 
Wide-field magnetometry imaging with a high spatial resolution provided by optical accessibility of the NV center is of particular interest for biology applications.
Previous reports on optical wide-field imaging have demonstrated potentials of novel imaging techniques using the NV center in diamond, such as spin noise imaging,\cite{Steinert2013} nuclear magnetic resonance (NMR) imaging,\cite{DeVience2015} electron spin resonance imaging,\cite{Simpson2017} and living cell imaging.\cite{LeSage2013, Glenn2015} 
Furthermore, the quantum behavior under ambient condition has a possibility for the realization of single-cell scale NMR analysis using \textit{in vivo} wide-field imaging.

The application of diamond wide-field magnetometry is limited by its magnetic sensitivity and frequency resolution, both of which are limited by the coherence time of the NV center.
The benchmark for biological applications is nanotesla sensitivity with subhertz resolution, which allows the observation of molecular structure identification from chemical shifts in NMR measurements.
For an alternating current (AC) field sensing, the XY8-based protocol with a coherence time ($T_2$) of the order of microseconds achieves a typical frequency resolution of around 10--\SI{100}{kHz} and a sensitivity of around \SIrtHz{100}{nT} using shallow NV center ensemble.\cite{Ishiwata2017, Tetienne2017a} 
Recently reported phase sensitive measurement protocol~\cite{Schmitt2017,Boss2017,Bucher2017} ``Qdyne'' drastically enhances the frequency resolution and the signal-to-noise ratio (SNR) without limitation of the coherence time imposed by the NV center.

For wide-field sensing, it is necessary to use a camera-type sensor, such as a charge-coupled device (CCD), which requires a long readout sequence compared with a high-speed detector such as an avalanche photo diode implemented for Qdyne in \citet{Schmitt2017}.
The detection time of an NV magnetometer is significantly shorter than the readout time; therefore, direct implementation of Qdyne leads to low sensitivity.
In this study, we propose a new phase sensitive protocol, ``iterative-Qdyne ({\iQdyne})'', suitable for wide-field sensing and demonstrate simultaneously both millihertz frequency resolution and nanotesla sensitivity, which are achieved by iterative measurements.

The Qdyne protocol utilizes the stroboscopic effect, which records the temporal evolution of the phase of the target signal by equally spaced XY8 measurements.
The external AC magnetic field causes of phase accumulation on the electron spin of NV center, which is dependent on the initial phase of the target signal, so the recorded phases stand for the time-evolution of the target signal.
Since the measurement interval $T_L$ is significantly longer than the target period $T_\mathrm{ac}$ due to the CCD readout time, the recorded phases are highly under-sampled.
Finally converting into the low-frequency region by the Fourier transform, its frequency resolution is limited by not the coherence time, but the total measurement time.
This leads to an ultrahigh frequency resolution up to the classical clock stability.
Fig.~\ref{fig:protocols}(a, upper) shows the directly implemented (without iteration, $N_\mathrm{rep}=1$) Qdyne protocol with CCD cameras that typically require a readout time of a few milliseconds, which is extraordinary longer than the XY8 detection time, which is typically around few microseconds.The ratio of the sensing time to the measurement period is crucial for the magnetic sensitivity.
In the case of slow detection, it is essential to enhance SNR and achieve a high sensitivity.
Fig.~\ref{fig:protocols}(a, lower) shows the proposed new protocol suitable for a slow detector we refer to as ``{\iQdyne}.'' 
The two followings are the key points: iterative XY8 measurements during a single exposure and adjustment of the sampling interval to an integer multiple of the target period (Fig.~\ref{fig:protocols}(b)).
The iteration increases the number of incident photons into the sensor, diminishes the shot-noise, and enhances the sensing duration ratio to the measurement period.
In order to effectively enhance SNR, it is favorable for the initial phases to be arranged identically amongst all XY8 measurements.
The maximum SNR is obtained when the interval of the XY8 measurements is adjusted to an integer multiple of the target period.
If a detuning exists, each XY8 measurement records different phases and the SNR is degenerated; that is, {\iQdyne} has a frequency dependence.
As described below, this bandwidth of {\iQdyne} is around the inverse of the sensing time duration.
Note that \citet{Boss2017} reports a type of the Qdyne method with the quantum nondemolition (QND) measurement of the order of milliseconds, but the QND measurement enhances SNR and plays a similar role as the iteration in our protocol.

In order to demonstrate the {\iQdyne} method, we used an ensemble of NV centers in a type-IIa$(100)$ diamond substrate.
The NV centers were fabricated by low-energy (\SI{6}{keV}) $\mathrm{N}^{+}$ ion implantation with a dose of \SI{2e13}{{cm}^{-2}} at an elevated temperature of \SI{600}{\degreeCelsius} and subsequent annealing for \SI{2}{h} at \SI{800}{\degreeCelsius}.
The surface density of the NV centers is \SI{1.8e12}{{cm}^{-2}} and the average coherence time ($T_2$) is approximately \SI{3}{\us}, which is estimated by a home-built confocal fluorescence microscope.
Fig.~\ref{fig:protocols}(c) shows a schematic diagram of the wide-field optical system.
A uniformly distributed excitation is implemented by a lens array and a \SI{100}{mW} laser (excitation wavelength: \SI{532}{nm}).
Fluorescence from the NV center is detected by a CCD camera (iXon3 860, Andor technology).
We define an observation area of \SI{20}{\um} square.
Microwaves for electron spin manipulation are applied via an $\mathrm{\Omega}$-shaped loop coil with a diameter of \SI{100}{\um}, and a radio frequency signal as a testing target is applied via a neighboring linear pattern.
In order to quantify the performance, we define the sensitivity $\eta$ as:
\begin{equation}
    \label{eq:sensitivity}
    \eta = \frac{b_z \sqrt{T_\mathrm{tot}}}{\mathrm{SNR}}
\end{equation}
where $b_z$, $T_\mathrm{tot}$ and $\mathrm{SNR}$ represent the applied testing field strength, total time of sensing, and SNR.
We refer ``S'' as the peak height and ``N'' as the standard deviation of residual differences between the measured spectrum and the estimated peak function (Fig.~\ref{fig:protocols}(d)).
We also define the frequency resolution as a frequency step.
Since we disciplined a digital timing generator and a function generator by a Rubidium oscillator (FS725, Stanford Research Systems), the linewidth of the test signal is very narrow, and its peak shape should be Sinc function because of spectral leakage.
Note that ``sensitivity'' in this paper is normalized as a sensitivity per \SI{1}{\um^2} area.

We set the test field period to $T_\mathrm{ac}=\SI{300}{ns}$ corresponding to \SI{3.333}{MHz} and detect by 48 pulsed XY8 measurements with sampling interval $T_S=\SI{13.5}{\us}$, so the ratio of $T_S / T_\mathrm{ac}$ is 45 (integer).
Fig.~\ref{fig:result_lockin}(a) shows directly implemented Qdyne in a wide-field and the {\iQdyne} protocol by 100 repetitive measurements during an exposure.
The data set consists of around 1000 points.
Owing to spectral leakage, it is possible to achieve higher SNR with shorter data set (Picket fence effect), so we find the highest SNR by varying the data length from 900 to 1000 points with each data set.
When the total measurement time is a multiple of the RF period, a delta-distributed spectrum is observed with no leakage.
The data is normalized by the total measurement time and the noise level.
Furthermore, the direct Qdyne data (red) is zoomed by 10 times for the vertical axis.
Direct Qdyne has a sensitivity of \SIrtHz{2.07(35)}{\micro T}.
Although its frequency resolution of \SI{336}{mHz} is much better than the coherence time ($T_2$) limited resolution of around 70 kHz, its sensitivity is limited by longer readout time compared with the detection time of the XY8 measurements.
In contrast, {\iQdyne} shows a sensitivity of \SIrtHz{65(2)}{nT} and a resolution of \SI{238}{mHz}.
The sensitivity of {\iQdyne} is substantially improved by iterative measurements.
We also measured the conventional XY8 sensitivity (data not shown), which was \SIrtHz{97(7)}{nT}, by triggering the RF field at every XY8 measurement.
{\iQdyne} provides comparable, but slightly better sensitivity.
This difference may be due to the amount of information: although the phase sensitive protocol detects the strength and the time evolution of the RF signal, the conventional XY8 technique detects only the RF field strength.
Fig.~\ref{fig:result_lockin}(b) shows the dependence of the sensitivity on the number of iterations during an exposure ($N_\mathrm{rep}$).
Assuming a shot-noise limited measurement, increasing $N_\mathrm{rep}$ will reduce the noise level with $\sqrt{N_\mathrm{rep}}$ scaling.
At the same time, increasing $N_\mathrm{rep}$ extends the total sensing time, and eventually we can ignore the readout time with $N_\mathrm{rep}\rightarrow\infty$.
Accordingly, we obtain the following relationship:
\begin{equation}
    \label{eq:sinsitivity_Nrep}
    \eta = \eta_\infty \sqrt{1 + \frac{T_\mathrm{read}}{T_S N_\mathrm{rep}}}
\end{equation}
where $T_\mathrm{read} \simeq \SI{3}{ms}$ represents the readout time of CCD and $\eta_\infty$ represents the optimal sensitivity (blue dashed line in Fig.~\ref{fig:result_lockin}(b)) to which the sensitivity will converge with sufficiently large $N_\mathrm{rep}$.
We obtain the optimal sensitivity $\eta_\infty = \SIrtHz{41}{nT}$ by numerical fit using Eq.~\ref{eq:sinsitivity_Nrep} for the measured sensitivities in the region that we can ignore the readout noise.
In the small-$N_\mathrm{rep}$ region, Eq.~\ref{eq:sinsitivity_Nrep} is far from the experimental data.
This is because this region is not shot-noise limited, but readout noise is dominant.
Note that an available sensitivity is limited by the finite quantum well depth of the sensor pixels.
In our case, the maximum iteration is around 1000 times and the sensitivity is \SIrtHz{48}{nT}.
The theoretical limit of the {\iQdyne} sensitivity has not been investigated, and further analysis of the Fisher information is required.

Fig.~\ref{fig:result_lockin}(c) shows simultaneous detection of two different RF signals separated by \SI{1}{Hz}.
The two peaks can be clearly distinguished because of the sharp frequency resolution of the {\iQdyne} protocol.
Fig.~\ref{fig:result_lockin}(d) shows the histogram of the estimated peak frequencies within the wide-field observation area.
We demonstrated that the peak positions are highly homogeneous.
The blue peak is taller and shaper than the green one.
This difference between the two distribution profiles (blue and green) may be due to the spectral leakage effect.
The fitting precision is affected by whether the RF frequencies on the frequency bin (no leakage) or between bins (leakage).

\begin{figure}
  \includegraphics[width=8.5cm]{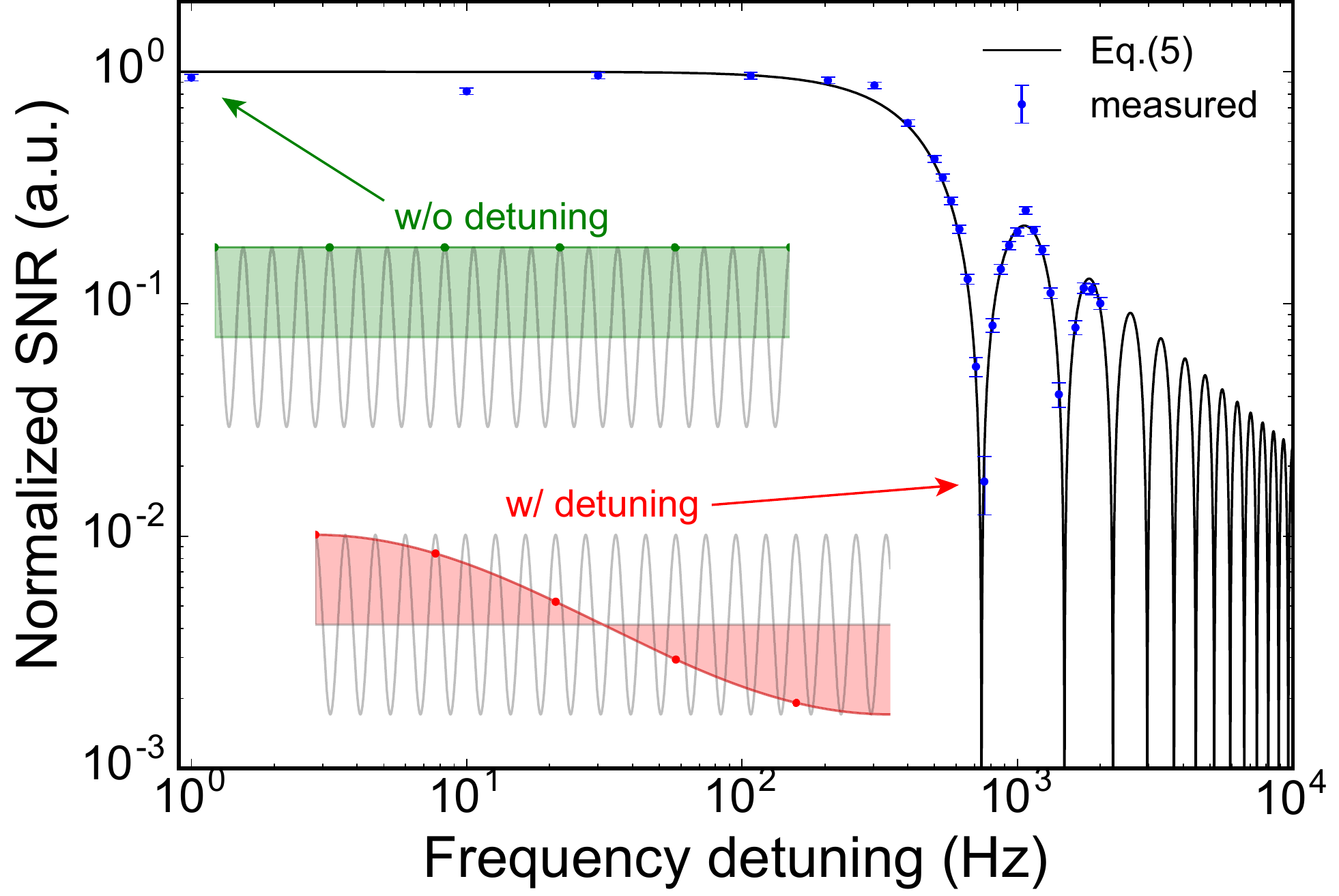}
  \caption{\label{fig:result_freq} Dependence of the SNR on the detuning of the test signal. The bandwidth as the first zero point is \SI{740}{Hz} corresponding to $(T_S N_\mathrm{rep})^{-1}$. Inset: (green) All XY8 measurements record the same RF signal phase, which leads to an enhanced SNR. (red) The recorded phases are different from each other, which leads to a low SNR. The error bars are estimated by the standard deviation of the fitting parameters.}
\end{figure}
As mentioned above, the RF detuning lowers the SNR.
In order to investigate the frequency response of {\iQdyne}, we measured the SNR with varying RF frequencies at $N_\mathrm{rep}=100$.
Fig.~\ref{fig:result_freq} shows the dependence of the SNR on the RF detuning.
It clearly shows insensitive dips at \SI{740}{Hz} and \SI{1.4}{kHz}, and the SNR revives around \SI{1.1}{kHz} and \SI{1.8}{kHz}.
The dips correspond to $(T_S N_\mathrm{rep})^{-1}$ and its multiples.
The RF phase advances by $\Delta \varphi = 2 \pi f_\mathrm{ac} \Delta T_S$ within the sampling interval $T_S$, where $\Delta T_S = T_S \mod T_\mathrm{ac}$ and $f_\mathrm{ac} = 1/T_\mathrm{ac}$ represents the sampling interval modulo the RF period and the frequency of the RF signal.
If the sampling interval is adjusted to a multiple of the RF period (no detuning), all the recorded signals are the same (Fig.~\ref{fig:result_freq}, green), and then the photon count in a single exposure $s$ will be enhanced by repeating.
However, nonzero detuning advances the RF phase during the sampling interval, and then the photon count $s$ does not increase repeatedly.
In particular, when the RF phase develops to $\pi$ during an exposure, the recorded phases cancel each other.
The photon counts will show no change in time, and the SNR falls to zero (Fig.~\ref{fig:result_freq}, red).
As the detuning increases further, accumulation of the phase advances revives and cancels again when the phase development is $2\pi$.
In this way, the frequency response shows a dip and revival.

Assuming that the RF field $b_z$ is weak and detuning $\Delta f_\mathrm{ac}$ is smaller than the bandwidth of XY8, each XY8 signal is proportional to $\cos \varphi$, where $\varphi$ is the initial RF phase at the front of the XY8 measurement.\cite{DeLange2011} 
Therefore, the result of a single exposure $s$ is the sum of $N_\mathrm{rep}$ times of the XY8 measurement with the advanced initial phase $\Delta \varphi$.
\begin{equation}
    \label{eq:s}
    s = \sum_{k=0}^{N_\mathrm{rep}-1} A b_z \cos \left( \phi + k \Delta \varphi \right)
\end{equation}
where $A$ represents a proportionality coefficient, and $\phi$ represents the initial RF phase of the first XY8 measurement.
After some algebra, we obtain 
\begin{equation}
    \label{eq:single_exposure}
    s = A b_z \cdot Z(\Delta f_\mathrm{ac}) \cdot \cos \left( \phi + \frac{N_\mathrm{rep}-1}{2} \Delta\varphi  \right)
\end{equation}
\begin{equation}
    \label{eq:Z}
    Z(\Delta f_\mathrm{ac}) = \frac{\sin(\pi N_\mathrm{rep} T_S \Delta f_\mathrm{ac})}{\sin(\pi T_S \Delta f_\mathrm{ac})}
\end{equation}
Here, $Z(\Delta f_\mathrm{ac})$ represents the frequency response of {\iQdyne} depicted as the solid line in Fig.~\ref{fig:result_freq}.
We define the bandwidth $W_{\mathit{i}\mathrm{Qdyne}}$ as the first zero point of the frequency response, 
\begin{equation}
    \label{eq:bandwidth}
    W_{\mathit{i}\mathrm{Qdyne}} = \frac{1}{N_\mathrm{rep} \cdot T_S}
\end{equation}
Therefore, the bandwidth of {\iQdyne} equals to the inverse of the detection time.

\begin{figure}
  \includegraphics[width=8.5cm]{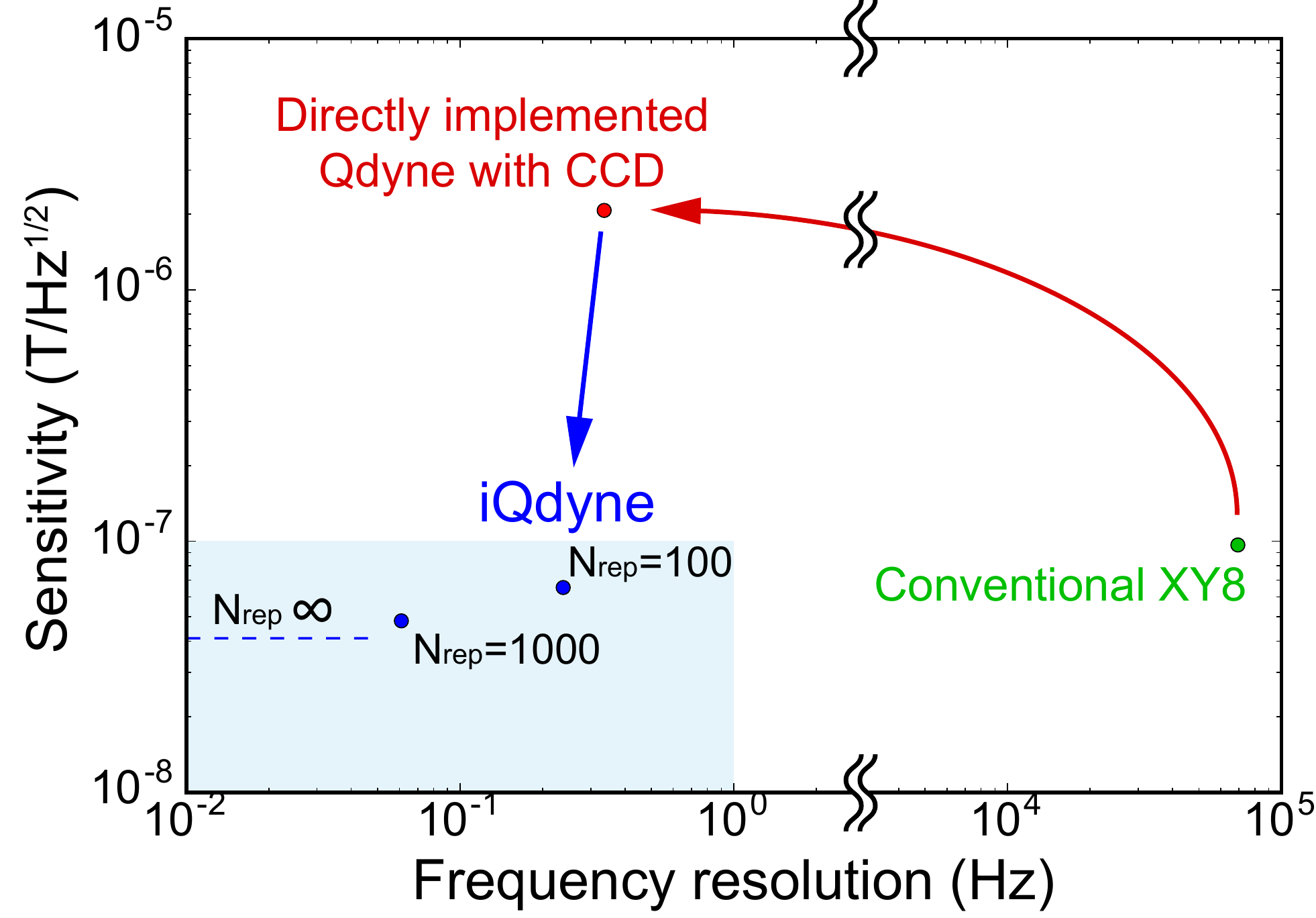}
  \caption{\label{fig:comparison_mapping} Magnetic sensitivity and frequency resolution of wide-field NV magnetometry: (green) conventional XY8 measurement, (red) directly implemented Qdyne in a wide-field, and (blue) proposed {\iQdyne} protocol. All dots were measured with the same setup.}
\end{figure}
Fig.~\ref{fig:comparison_mapping} shows the performance comparison among the AC sensing protocols.
The conventional XY8 technique (green) provides high sensitivity, but its resolution is of the order of kilohertz.
Previous reports using Hahn echo\cite{Pham2011} and dynamical decoupling\cite{DeVience2015} were in a similar performance region, because both the sensitivity and resolution were limited by the coherence time.
The Qdyne protocol liberate the frequency resolution from the coherence limit and enhances the resolution up to five orders of magnitude, but in a wide-field, the low duty ratio of the sensing time to the total measurement time deteriorates the sensitivity (red arrow).
In contrast, our proposed {\iQdyne} protocol improves the sensitivity by 40 times with the sharp frequency resolution (blue arrow).
Considering nanoscale NMR detection, since the spin noise is around \SI{300}{nT}-rms at \SI{10}{nm} below the surface\cite{Pham2016}, nanotesla sensitivity allows us to recognize it.
However, the requirement for the frequency resolution is quite severe because the chemical shifts in an NMR signal appear as a few hertz frequency variance.
Our protocol accommodates both the sensitivity and resolution (blue filled area); thus, it can be a promising candidate for NMR imaging.

In summary, this study proposes a new phase sensitive protocol {\iQdyne} that can be used with low speed detectors such as a CCD camera.
The {\iQdyne} greatly enhances the magnetic sensitivity and the frequency resolution by iterative measurements.
We demonstrated wide-field magnetometry with a resolution of \SI{238}{mHz} and a sensitivity of \SIrtHz{65}{nT} per \SI{1}{\um^2} with 100 iterations.
This method allows us to distinguish proximate two frequencies, which are unresolvable by the conventional XY8 technique.
We confirmed experimentally and analytically that the frequency window of {\iQdyne} was the inverse of the total sensing time.
The {\iQdyne} offers a route for quantum magnetometry applications for biological measurements such as cellular bio-magnetism imaging and wide-field NMR imaging.

We thank T. Makino from AIST for fabricating our diamond sample.
This work was supported by JST CREST Grant Number JPMJCR1333, Japan.

\bibliography{bib_mizuno1.bib}

\end{document}